\documentclass[letterpaper]{article} % DO NOT CHANGE THIS
\usepackage{aaai25}  % DO NOT CHANGE THIS
\usepackage{times}  % DO NOT CHANGE THIS
\usepackage{helvet}  % DO NOT CHANGE THIS
\usepackage{courier}  % DO NOT CHANGE THIS
\usepackage[hyphens]{url}  % DO NOT CHANGE THIS
\usepackage{graphicx} % DO NOT CHANGE THIS
\usepackage{natbib}  % DO NOT CHANGE THIS AND DO NOT ADD ANY OPTIONS TO IT
\usepackage{caption} % DO NOT CHANGE THIS AND DO NOT ADD ANY OPTIONS TO IT
\usepackage{algorithm}
\usepackage{algorithmic}

\usepackage{amsthm,amsmath,amssymb}
\usepackage{cite}
\usepackage{booktabs}
\usepackage{multirow}
\usepackage{subfigure}
\usepackage[misc]{ifsym}
\usepackage{xcolor}
\usepackage{pifont}

\usepackage{newfloat}
\usepackage{listings}
\DeclareCaptionStyle{ruled}{labelfont=normalfont,labelsep=colon,strut=off} % DO NOT CHANGE THIS
\floatstyle{ruled}
\newfloat{listing}{tb}{lst}{}
\floatname{listing}{Listing}
\title{Augmenting Sequential Recommendation with Balanced Relevance and Diversity}
\author {
    Yizhou Dang \textsuperscript{\rm 1},
    Jiahui Zhang \textsuperscript{\rm 1},
    Yuting Liu \textsuperscript{\rm 1},
    Enneng Yang \textsuperscript{\rm 1}, 
    Yuliang Liang \textsuperscript{\rm 1}\\
    Guibing Guo \textsuperscript{\rm 1}\thanks{Corresponding Authors.},
    Jianzhe Zhao \textsuperscript{\rm 1}$\textsuperscript{\rm *}$,
    Xingwei Wang \textsuperscript{\rm 2}
}
\affiliations {
    \textsuperscript{\rm 1} Software College, Northeastern University, China \\
    \textsuperscript{\rm 2} School of Computer Science and Engineering, Northeastern University, China\\
    \{yzdang,yutingliu,ennengyang,liangyuliang\}@stumail.neu.edu.cn\\ \{guogb,zhaojz\}@swc.neu.edu.cn,  \{zhangjiahui,wangxw\}@mail.neu.edu.cn
}

\begin{document}

\maketitle

\begin{abstract}
By generating new yet effective data, data augmentation has become a promising method to mitigate the data sparsity problem in sequential recommendation. Existing works focus on augmenting the original data but rarely explore the issue of imbalanced relevance and diversity for augmented data, leading to semantic drift problems or limited performance improvements. In this paper, we propose a novel \underline{B}alanced data \underline{A}ugmentation Plugin for \underline{S}equential \underline{Rec}ommendation (BASRec) to generate data that balance relevance and diversity. BASRec consists of two modules: Single-sequence Augmentation and Cross-sequence Augmentation. The former leverages the randomness of the heuristic operators to generate diverse sequences for a single user, after which the diverse and the original sequences are fused at the representation level to obtain relevance. Further, we devise a reweighting strategy to enable the model to learn the preferences based on the two properties adaptively. The Cross-sequence Augmentation performs nonlinear mixing between different sequence representations from two directions. It produces virtual sequence representations that are diverse enough but retain the vital semantics of the original sequences. These two modules enhance the model to discover fine-grained preferences knowledge from single-user and cross-user perspectives. Extensive experiments verify the effectiveness of BASRec. The average improvement is up to 72.0\% on GRU4Rec, 33.8\% on SASRec, and 68.5\% on FMLP-Rec. We demonstrate that BASRec generates data with a better balance between relevance and diversity than existing methods. The source code is available at \url{https://github.com/KingGugu/BASRec}.
\end{abstract}

\section{Introduction}
As an essential branch of recommender systems, sequential recommendation (SR) has received much attention due to its well-consistency with real-world recommendation situations. However, the widespread problem of data sparsity limits the SR model's performance \cite{jing2023contrastive}. For this reason, researchers have proposed many data augmentation methods to mitigate this phenomenon \cite{dang2023uniform, dang2024repeated}. Earlier work used heuristic methods to directly augment sequences and mix them to the training process, such as Sliding Windows \cite{tang2018personalized} and Dropout \cite{tan2016improved}. Later, some researchers generated high-quality augmented data by counterfactual thinking \cite{wang2021counterfactual}, diffusion models \cite{liu2023diffusion} or bi-directional transformer \cite{jiang2021sequential}. With the success of self-supervised learning \cite{chen2020simple}, many sequence-level augmentation operators have been proposed for contrastive learning \cite{liu2021contrastive, xie2022contrastive}.

\begin{figure}[!t]
	\centering
	\includegraphics[scale=0.60]{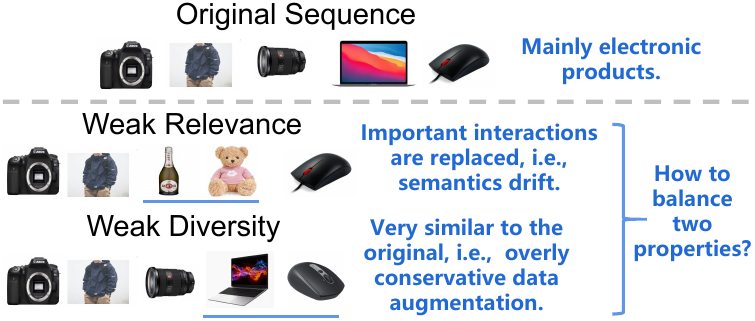}
	\caption{An illustration of imbalanced relevance and diversity issues in sequential data augmentation.}
	\label{fig:example}
	\vspace{-2mm}
\end{figure}

Despite the effectiveness, the imbalance between relevance and diversity for augmented data remains to be solved \cite{bian2022relevant}. `Relevance' means that the augmented data should have transition patterns similar to the original data to avoid semantics drift problems. `Diversity' means that the augmented data should contain sufficient variations to enable the model to explore more user preferences knowledge and improve its performance. However, these two factors are often conflicting and challenging to trade off. For example, as shown in Figure \ref{fig:example}, heuristic augmentation and operators proposed in recent years perform cropping, masking, substitution, or shuffle to the original sequence. The new sequences may deviate from the original data, resulting in weak correlations. Model- and representation-level augmentation methods, while improving the correlation of the generated data through well-designed generation modules, fail to produce sufficient diversity. These conservative augmented data make it challenging to improve the model performance further. Intuitively, we can obtain suitable samples by directly merging the two augmented sequences in Figure \ref{fig:example}. However, directly merging them will corrupt the original sequence patterns, and irrelevant or repeated items will be retained, resulting in inaccurate representation learning. In addition, existing methods focus on single-sequence augmentation and lack consideration for the cross-sequence preference patterns and semantic information.

In this paper, inspired by previous work based on mixup \cite{zhang2018mixup, bian2022relevant}, we propose BASRec, \underline{B}alanced data \underline{A}ugmentation plugin for \underline{S}equential \underline{Rec}ommendation. Our core idea is to generate new samples that balance relevance and diversity through representation-level fusion. The BASRec consists of two modules: Single-sequence Augmentation and Cross-sequence Augmentation. For Single-sequence Augmentation, we propose two new mixup-based data augmentation operators, M-Substitute and M-Reorder. On top of the traditional operator, we improve the diversity by sampling the operation weights from a uniform distribution, while the correlation is obtained by fusing the original and the augmented sequence representation. Further, we adaptively reweight the training loss with the operators' augmentation weights and mixup weights, allowing the model to learn based on the difference between the augmented and the original samples. So far, all augmentations are limited to a single sequence. For Cross-sequence Augmentation, we propose to explore the semantics of sequence and preferences knowledge among different users. The traditional mixup operation can only generate samples in the linear dimension for a pair of samples \cite{guo2020nonlinear}. In order to further improve the space of synthetic samples and enable the model to discover the fine-grained knowledge among different users, we adopt a nonlinear mixup approach for fusion. However, assigning a weight to each parameter introduces a significant computational overhead. We decompose this process and further present the Item-wise and Feature-wise mixup. Our method generates new training samples in the representation space that relate to the original data but contain diverse transition patterns, balancing the two properties. Finally, we employ a two-stage learning strategy to avoid the noise and convergence instability introduced by the hybrid representation at the beginning of training. 

Extensive experiments on four real-world datasets with four base SR models demonstrate that our proposed BASRec achieves significant improvements. We compare BASRec with heuristic and training-required data augmentation methods to further validate its superiority. Besides, We experimentally demonstrate that the data generated by BASRec strikes a better balance between the two properties. The main contributions can be summarized follows:
\begin{itemize}
    \item We emphasize the unbalanced relevance and diversity of current data augmentation methods and propose a Balanced Data Augmentation Plugin for Sequential Recommendation, which can be seamlessly integrated into most existing sequential recommendation models.
    \item We design two key modules, Single-sequence Augmentation and Cross-sequence Augmentation. They perform augmentation and fusion operations to synthesize new samples that balance relevance and diversity.
    \item We conduct comprehensive experiments on real-world datasets to demonstrate the effectiveness of BASRec, showing significant improvements over various sequential models. Our approach also achieves competitive performance compared to the data augmentation baselines.
\end{itemize}

\section{Related Work}

\subsection{Sequential Recommendation}
Sequential recommendation aims to provide users with personalized suggestions based on their historical behaviors. Early works leveraged Markov Chains \cite{rendle2010factorizing} and session-based KNN \cite{he2016fusing, hu2020modeling} to model sequential data. Later, researchers adopted CNNs \cite{tang2018personalized} and RNNs \cite{liu2016context} to capture the relationships among items. For instance, NextItNet \cite{yuan2019simple} combined masked filters with 1D dilated convolutions to model the sequential dependencies. GRU4Rec \cite{hidasi2015session} used gated recurrent units to capture behavioral patterns. More recently, Transformers \cite{vaswani2017attention} showed extraordinary performance in learning item importance and behavior relevance for next-item prediction. SASRec \cite{kang2018self} is the representative work that adopted the multi-head attention mechanism to perform sequential recommendation. BERT4Rec \cite{sun2019bert4rec} introduced a cloze task with a bidirectional attentive encoder. To model dynamic uncertainty and capture collaborative transitivity, STOSA \cite{fan2022sequential} embeds each item as a stochastic Gaussian distribution and devises a Wasserstein Self-Attention module to characterize item-item relationships. In addition, cross-domain \cite{zhao2023sequential}, multi-modal \cite{zhang2023multimodal}, and multi-behavior \cite{su2023personalized} SR tasks have also been widely explored.

\subsection{Data Augmentation for SR}
Data augmentation has been widely used in many domains to improve model performance and robustness \cite{dang2024data}. It also received extensive attention in sequential recommendation. Slide Windows \cite{tang2018personalized} and Dropout \cite{tan2016improved} are pioneer works that split one sequence into many sub-sequences or discard some items from the original data. Later, many training-required data synthesis methods are proposed since heuristics may produce low-quality augmented data. For example, DiffASR \cite{liu2023diffusion} adopted the diffusion model for sequence generation. Two guide strategies are designed to control the model to generate the items corresponding to the raw data. CASR \cite{wang2021counterfactual} proposed substituting some previously purchased items with other unknown items based on counterfactual thinking. L2Aug \cite{wang2022learning} enhanced casual users' sequences by learning from core users' interaction patterns. Besides, some work has explored using self-supervised signals to mitigate data sparsity in SR. CL4SRec \cite{xie2022contrastive} constructed contrastive views by three augmentation operators, i.e., item crop, mask, and reorder. DuoRec \cite{qiu2022contrastive} explored the representation degeneration issue in SR and solved it by contrastive regularization. ReDA \cite{bian2022relevant} adopted a neural retriever to retrieve augmentation users and conducted two types of representation augmentation. 

Unlike these studies, we propose a Balanced Data Augmentation Plugin to generate data that considers relevance and diversity. It performs augmentation by elaborate augmentation and mixup operations in the representation space.

\section{Preliminaries}

\subsection{Problem Formulation}
Suppose we have user and item sets denoted by $\mathcal{U}$ and $\mathcal{V}$, respectively. Each user $u \in \mathcal{U}$ is associated with a sequence of interacted items in chronological order $s_u=[v_1, v_2, \ldots, v_{\left|s_u\right|}]$, where $v_j \in \mathcal{V}$ indicate the item that user $u$ has interacted with at time step $j$. The $\left|s_u\right|$ is the sequence length. Given the sequences of interacted items $s_u$, SR aims to accurately predict the possible item $v^{*}$ that user $u$ will interact with at time step $\left|s_u\right|+1$, formulated as follows:
\begin{equation}
    \underset{v^{*} \in \mathcal{V}}{\arg \max} \;\; P\left(v_{\left|s_u\right|+1}=v^{*} \mid s_u \right).
\end{equation}
The model will calculate the probability of all candidate items and select the highest one for recommendation.

\subsection{Mixup for Data Augmentation}
Mixup \cite{zhang2018mixup, zhang2020seqmix} is a simple yet effective data augmentation method. It implements linear interpolation in the input space to construct virtual training data. Given two input samples $x_i, x_j$ along with the labels $y_i, y_j$, the mixup process can be formulated as:
\begin{equation}
\tilde{x}=\lambda \cdot x_i+(1-\lambda) \cdot x_j,
\end{equation}
\begin{equation}
\tilde{y}=\lambda \cdot y_i+ (1-\lambda) \cdot y_j.
\end{equation}
where $\lambda \sim \operatorname{Beta}(\alpha, \alpha)$ is the mixup coefficient from beta distribution. Some work in the recommendation field has explored the use of mixup to synthesize hard negative samples for training graph neural networks better \cite{huang2021mixgcf} or to improve the representation of tail sessions \cite{yang2023loam}. However, these efforts failed to balance relevance and diversity when performing augmentation. They may generate harmful or overly conservative augmented data. Also, they are limited by the type of backbone network and have lower generalization capability \cite{bian2022relevant}.

\section{Ours: BASRec}
\begin{figure*}[!t]
	\centering
	\includegraphics[scale=0.43]{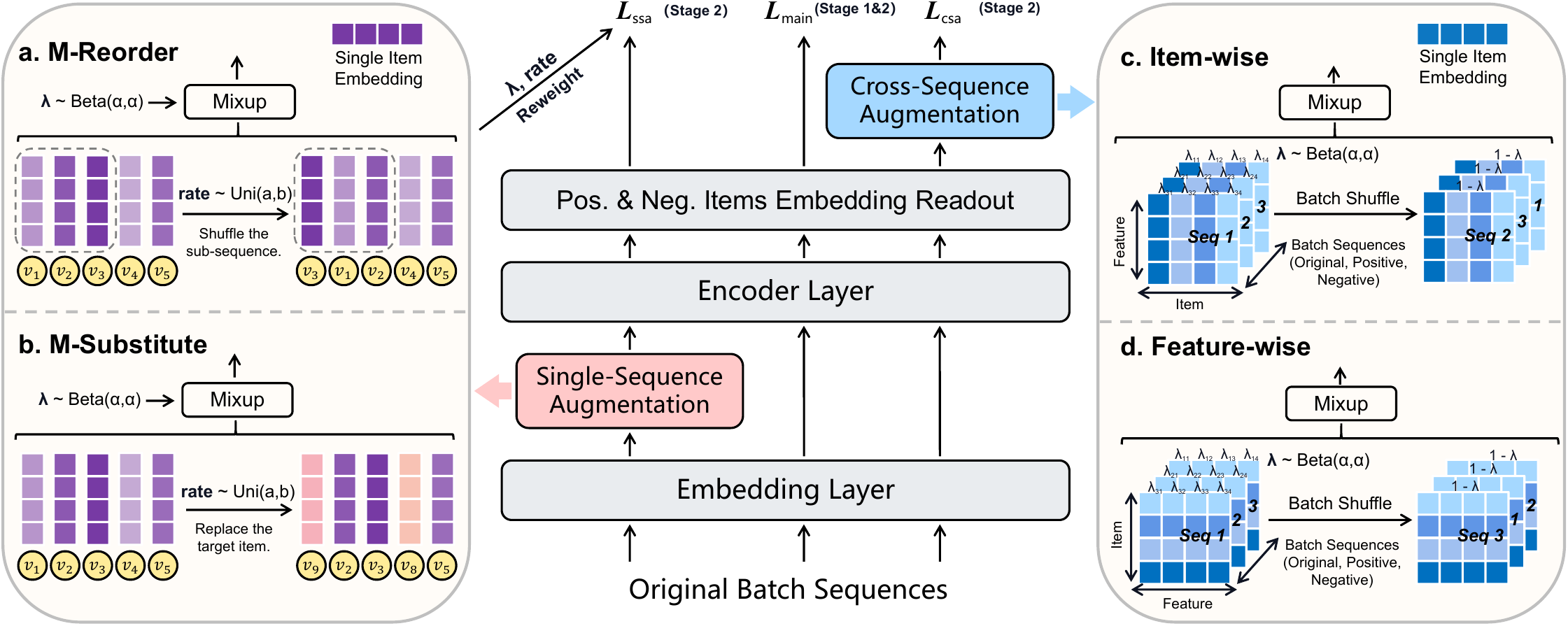}
	\caption{The overview of the proposed Balanced data Augmentation method.}
	\label{fig:framework}
	\vspace{-2mm}
\end{figure*}

In this section, we present our proposed BASRec. The overall framework is illustrated in Figure \ref{fig:framework}. Our approach consists of two separate modules: Single-sequence Augmentation and Cross-sequence Augmentation. After that, we introduce the training and inference process of BASRec. Finally, we provide a discussion about our and existing methods. 

\subsection{Single-sequence Augmentation}
Single-sequence Augmentation generates new samples by mixing the representations of items in the original sequence with those in the augmented sequence. Specifically, we propose two new operators, M-Reorder and M-Substitute, to accomplish this augmentation operation.

The standard SR paradigm \cite{kang2018self, bian2022relevant} maintains an item embedding matrix $\mathbf{M}_\mathcal{V} \in \mathbb{R}^{|\mathcal{V}| \times D}$. The matrix project the high-dimensional one-hot representation of an item to low-dimensional dense representations. Given an original user sequence $s_u=[v_1, v_2, \ldots, v_{\left|s_u\right|}]$, the $\operatorname{Look-up}$ operation will be applied for $\mathbf{M}_\mathcal{V}$ to get a sequence of item representation, i.e., $E_u=[m_{v_1}, m_{v_2}, \ldots, m_{v_{\left|s_u\right|}}]$. Note that we omit the padding for short sequences and intercepting for long sequences.

\ \\ \noindent \textbf{M-Reorder.} Given an original sequence $s_u$, M-Reorder first selects a sub-sequence with length $c=rate \cdot \left|s_u\right|$. Unlike traditional operators that preform augmentation with a fixed $rate$, we further extend the augmentation possibilities by drawing $rate$ from a uniform distribution:
\begin{equation}
rate \sim \operatorname{Uniform}(a, b),
\label{eq:rate}
\end{equation}
where $a$ and $b$ are hyper-parameters and $0$ \textless $a$ \textless $b$ \textless $1$. Then, we randomly shuffle this sub-sequence $\left[v_i, \cdots, v_{i+c-1}\right]$ as $\left[v_i^{\prime}, \ldots, v_{i+c-1}^{\prime}\right]$ and get the augmented item sequence $s_u^{\prime}$: 
\begin{equation}
s_u^{\prime}=\operatorname{Reorder}\left(s_u\right)=\left[v_1, v_2, \cdots, v_i^{\prime}, \cdots, v_{i+r-1}^{\prime}, \cdots, v_n\right].
\end{equation}
Unlike traditional operators that directly use $s_u^{\prime}$ as a new sample to participate in model training, we mix up the sequence of item representation corresponding to $s_u$ and $s_u^{\prime}$ to generate new training samples in the representation space:
\begin{equation}
E_u^{\prime} = \operatorname{Look-up}\left(\mathbf{M}_\mathcal{V}, s_u^{\prime}\right),
\label{eq:reorder-look-up}
\end{equation}
\begin{equation}
E_u^{In} = \lambda \cdot E_u + (1-\lambda) \cdot E_u^{\prime},
\label{eq:reorder-mixup}
\end{equation}
where $\lambda \sim \operatorname{Beta}(\alpha, \alpha)$ is the mixup weight and $E_u^{In}$ is augment representation used for model training.

\ \\ \noindent \textbf{M-Substitute.} This operator is similar to M-Reorder. Given an original sequence $s_u$, it first randomly selects $c=rate \cdot \left|s_u\right|$ different indices $\left\{\operatorname{idx}_1, \operatorname{idx}_2, \ldots, \operatorname{idx}_c\right\}$, where $rate$ is sampled following Eq. \ref{eq:rate}. Then, we replace each with a correlated item based on the selected indices. We adopt the cosine similarity method as \cite{liu2021contrastive} to select the items to substitute. The above process can be formulated as:
\begin{equation}
s_u^{\prime}=\operatorname{Substitute}\left(s_u\right)=\left[v_1, v_2, \ldots, \bar{v}_{\mathrm{idx}_i}, \ldots, v_{\left|s_u\right|}\right].
\end{equation}
Following the same steps as in M-Reorder, through Eq. \ref{eq:reorder-look-up} and Eq. \ref{eq:reorder-mixup}, we can obtain the augmented representation $E_u^{In}$.

\ \\ \noindent \textbf{Adaptive Loss Weighting.} 
The augmentation of the original representation by the two operators comes from two main dimensions: 1) The change of the operator to the original interaction sequence, i.e., the $rate$ of operators. 2) The mixing of the new sequence representation with the original one, i.e., the $\lambda$ of mixup. To further measure this augmentation process so that the model can adaptively learn the preferences based on the relevance and diversity of the augmentation, we propose an adaptive loss weighting strategy. 
Inspired by previous work \cite{yang2023debiased}, based on the $rate$ from two operators and mixup coefficient $\lambda$ from $\operatorname{Beta}(\alpha, \alpha)$, we define the transformation as follows:
\begin{equation}
\omega^{(1)}= 1 / \left(rate \cdot \lambda \right); \: \omega^{(2)}=\frac{\omega^{(1)}-\omega_{\min}^{(1)}}{\omega_{\max}^{(1)}-\omega_{\min}^{(1)}}.
\end{equation}
We adopt $\omega = \omega^{(2)}$ as the output. For each augmented representation $E_u^{f}$, we assign it with an exclusive weight. This weighting process also guides the model in distinguishing how much of the original representation has been injected with the new representation, further improving the robustness of the model. This weight will be used in model training, and we will introduce it in the Model Training section.

\subsection{Cross-sequence Augmentation}
In Single-sequence Augmentation, the newly generated samples are limited to only independent single users. In recommender systems, each user is likely to have overlapping behaviors and preferences with other users, i.e., collaboration information \cite{luo2024collaborative, cheng2024empowering}. Therefore, for Cross-sequence Augmentation, we generate new samples by mixing the different sequence output to discover the preference knowledge among different users further. By feeding the sequence representation $E_u$ and $E_u^{In}$ into the Encoder, we can obtain the output representation of the sequence $H_u$ and $H_u^{In}$. In sequential recommendation, user representations are usually modeled implicitly, so $H_u$ and $H_u^{In}$ can also be interpreted as user representations. Note that the two modules we proposed are independent augmentation lines. The representation of a sequence that has been augmented by Single-sequence Augmentation will not be fed into Cross-sequence Augmentation.

The base mixup can only create samples in linear space. Inspired by the nonlinear mixup \cite{guo2020nonlinear}, we attempt to assign different weights to every parameter in $H_u$. However, this mixup strategy incurs significant computational overhead. Thus, we further decompose this process into Item-wise mixup and Feature-wise mixup. Given a batch of sequences $\left\{s_u\right\}_{u=1}^B$, we can obtain the corresponding batch of representations $\left\{H_u\right\}_{u=1}^B \in R^{B \times N \times D}$, where $B$, $N$ and $D$ are the batch size, maximum sequence length, and embedding dimensions, respectively. We start by shuffling $\left\{H_u\right\}_{u=1}^B$ from the batch perspective:
\begin{equation}
\left\{H_u^{\prime}\right\}_{u=1}^B = \operatorname{Shuffle} \left\{H_u\right\}_{u=1}^B.
\end{equation}
For \textbf{Item-wise Nonlinear Mixup}, we sample a mixup weights matrix $\Lambda_I \in R^{B \times N}$ from the $\operatorname{Beta}(\alpha, \alpha)$, where $N$ is the predefined maximum sequence length. The mixup is performed as follows:
\begin{equation}
\left\{H_u^{Out} \right\}_{u=1}^B = \Lambda_I \circ \left\{H_u\right\}_{u=1}^B + (1-\Lambda_I) \circ \left\{H_u^{\prime}\right\}_{u=1}^B,
\end{equation}
where $\circ$ denotes the Hadamard product. For \textbf{Feature-wise Nonlinear Mixup}, we sample a mixup weights matrix $\Lambda_F \in R^{B \times D}$ from the $\operatorname{Beta}(\alpha, \alpha)$ and perform  similar operation:
\begin{equation}
\left\{H_u^{Out} \right\}_{u=1}^B = \Lambda_F \circ \left\{H_u\right\}_{u=1}^B + (1-\Lambda_F) \circ \left\{H_u^{\prime}\right\}_{u=1}^B.
\end{equation}

The Cross-sequence Augmentation will simultaneously perform the same mixup process for the representations of positive and negative items. The final output is used to calculate the recommendation loss. The above two processes can be executed multiple times, and each mixing produces new virtual representations in the representation space. 

\subsection{Model Training}
Following previous works \cite{kang2018self,dang2023ticoserec}, we adopt the commonly used binary cross-entropy (BCE) loss for the sequential recommendation task:
\begin{equation}
\begin{aligned}
\mathcal{L}_{main} & = \operatorname{BCE}\left(H_u, E_u^{+}, E_u^{-}\right) \\ & = 
- \left[\log \left(\sigma\left(H_u \cdot E_u^{+}\right)\right)+\log \left(1 -\sigma\left(H_u \cdot E_u^{-}\right)\right)\right],
\end{aligned}
\label{eq:main}
\end{equation}
where $H_u, E_u^{+}$ and $E_u^{-}$ denote the representations of the user, the positive and negative items, respectively. $\sigma()$ is the sigmoid function. The representations involved in the $\mathcal{L}_{main}$ will not be augmented. For each sequence, we apply two operators for augmentation, respectively. The loss is calculated based on $H_u^{In}$, original positive and negative item representations. Further, we use the pre-computed weights $\omega$ to reweight this objective function:
\begin{equation}
\mathcal{L}_{ssa} = \omega \cdot \operatorname{BCE} \left(H_u^{In}, E_u^{+}, E_u^{-}\right).
\end{equation}
For Cross-sequence Augmentation, we use all mixed representations to calculate the recommendation loss:
\begin{equation}
\mathcal{L}_{csa} = \operatorname{BCE} \left(H_u^{Out}, E_u^{Out+}, E_u^{Out-}\right).
\end{equation}

Mixing representations at the beginning of training with Randomly initialized embedding may introduce noise and convergence problems. To tackle this, we adopt a two-stage training strategy. In the first stage, we follow the standard sequential recommendation model training process, with the primary objective of facilitating the learning of high-quality representations for items. We only use Eq. \ref{eq:main} as the objective function at this stage. In the second stage, we employ the two augmentation modules as described previously:
\begin{equation}
\mathcal{L} = \mathcal{L}_{main} + \mathcal{L}_{ssa} + \mathcal{L}_{csa}.
\label{eq:main+aug}
\end{equation}
We do not set additional loss weights for $\mathcal{L}_{ssa}$ and $\mathcal{L}_{csa}$ for the following reasons: 1) The $\mathcal{L}_{ssa}$ has been reweighted in the previous sections. 2) For the two proposed modules, we treat the data augmented by them equally to the original data during the training process. Each augmented data can be considered as a new user and interaction generated in the representation space. During the inference phase, all augmentation modules are deactivated.

\subsection{Discussion and Analysis}
\noindent \textbf{Comparison with Existing Methods.} 
For Single-sequence Augmentation, sampling $rate$ from a uniform distribution improves the diversity of the augmented sequence. The fusion with the original sequence representation and the loss reweighting process enables the model to learn based on the correlation of the augmented samples with the original samples. For Cross-sequence Augmentation, our cross-user nonlinear mixup strategy endows the learning process with diverse but relevant preference knowledge and collaborative signals among different users. Traditional operators \cite{liu2021contrastive, xie2022contrastive} that directly edit the original sequence with a fixed $rate$ may lead to problems of conservative data augmentation or semantics drift. Editing the original sequence may also remove critical interactions \cite{dang2023uniform}. Some work craft trainable data generators to augment the data \cite{wang2021counterfactual, liu2023diffusion, wang2022learning}. However, these methods can only produce discrete user interactions while introducing additional learnable parameters. Our method can generate more samples at the sequence level or across sequences in the representation space without additional model parameters. Besides, some approaches are also limited by the type of backbone network \cite{jiang2021sequential, bian2022relevant}, whereas BASRec is a model-agnostic augmentation plugin. 

\ \\ \noindent \textbf{Complexity Analysis.} We choose SASRec as the backbone model for explanation. Other choices can be analyzed similarly. Since Our BASRec does not introduce any auxiliary learnable parameters, the model size of BASRec is identical to SASRec. The time complexity of SASRec is mainly due to the self-attention module, which is $O\left(N^2 D |\mathcal{U}| \right)$ \cite{xie2022contrastive}. The time complexity for calculating loss is $\mathcal{O}\left(N D|\mathcal{U}|\right)$. Considering our method, for two operators in BASRec, the complexity of operation is $O\left((a+b) N |\mathcal{U}|/2 \right)$. Suppose we need to perform a total of $Q$ mixup operations for each sequence, so the time complexity is $O\left(Q D |\mathcal{U}|\right)$ since the mixing process is performed through Hadamard product. The total time complexity of SASRec and BASRec are $\mathcal{O}\left(\left(N^2 + N\right) D|\mathcal{U}|\right)$ and $\mathcal{O}\left(\left(N^2 + N + Q\right) D|\mathcal{U}|\right)$\footnote{The time cost of operators, $O\left((a+b) N |\mathcal{U}|/2 \right)$, is omitted since it has a lower complexity order.}, respectively. Their analytical complexity is the same in magnitude. Our BASRec can generate data with acceptable additional time costs.

\begin{table}[!t]
	\centering
	\scalebox{0.96}{
		\begin{tabular}{l|rrrr}
			\toprule \textbf{{Dataset}} & \textbf{Beauty} &\textbf{Sports} & \textbf{Yelp} & \textbf{Home} \\
			\midrule 
		   {\# Users} & 22,363 & 35,958 & 30,431 & 66,519 \\
			 {\# Items} & 12,101 & 18,357 & 20,033 & 28,237 \\
			 {\# Inter} & 198,502 & 296,337 & 316,354 & 551,682 \\
              {\# AvgLen} & 8.9 & 8.3 & 10.4 & 8.3 \\
			 {Sparsity} & 99.92\% & 99.95\% & 99.95\% & 99.97\% \\
			\bottomrule
	\end{tabular}}
  	\caption{The statistics of four datasets. The `Inter' and `AvgLen' denote the number of interactions and average length.}
	\label{tab:datasets}
 \vspace{-1em}
\end{table}

\section{Experiments}

\subsection{Experimental Settings}
\noindent \textbf{Datasets.} We adopt four widely-used public datasets: Beauty, Sports, and Home are obtained from Amazon \cite{Amazon} with user reviews of products. Yelp\footnote{https://www.yelp.com/dataset} is a business dataset. We use the transaction records after January 1st, 2019. Users/items with fewer than five interactions are filtered out \cite{liu2021contrastive}. The detailed statistics are summarized in Table \ref{tab:datasets}.

\ \\ \noindent \textbf{Baselines.} The baselines consist of three categories. The first category is general models to validate the effectiveness of BASRec, including GRU4Rec \cite{hidasi2015session}, NextItNet \cite{yuan2019simple}, SASRec \cite{kang2018self} and FMLPRec \cite{zhou2022filter}. These models employ diverse architectures, including RNN, CNN, Transformer, and MLP. The second category is heuristic augmentation methods: Random (Ran) and Random-seq (Ran-S) \cite{liu2023diffusion}, Slide Windows (SW) \cite{tang2018personalized}, CMR \cite{xie2022contrastive} and CMRSI \cite{liu2021contrastive}. The third category is training-required augmentation models: ASReP \cite{jiang2021sequential}, DiffuASR \cite{liu2023diffusion}, and CL4SRec \cite{xie2022contrastive}. Details about baselines are provided in the Appendix. We do not include methods ReDA \cite{bian2022relevant}, CASR \cite{wang2021counterfactual}, and L2Aug \cite{wang2022learning} since they do not provide open-source codes for reliable reproduction (empty code repository or no available code repository links).

\ \\ \noindent \textbf{Implementation Details.} For all baselines, we adopt the implementation provided by the authors. We set the embedding size to 64 and the batch size to 256. The maximum sequence length is set to 50. To ensure fair comparisons, we carefully set and tune all other hyper-parameters of each method as reported and suggested in the original papers. We use the Adam \cite{2014Adam} optimizer with the learning rate 0.001, $\beta_1=0.9$, $\beta_2=0.999$. For BASRec, we tune the $\alpha, a, b$ in the range of $\{0.2,0.3,0.4,0.5,0.6\}$, $\{0.1,0.2,0.3\}$, $\{0.6,0.7,0.8\}$, respectively. We conduct five runs and report the average results for all methods. Generally, \emph{greater} values imply \emph{better} ranking accuracy.

\setlength{\tabcolsep}{1mm}{
\begin{table*}[!t]
  \centering
	\scalebox{0.77}{
    \begin{tabular}{c|cccc|cccc|cccc|cccc}
    \toprule
    \multirow{2}[2]{*}{Method} & \multicolumn{4}{c|}{Beauty} & \multicolumn{4}{c|}{Sports} & \multicolumn{4}{c|}{Yelp} & \multicolumn{4}{c}{Home} \\
  & N@10 & H@10 & N@20 & H@20 & N@10 & H@10 & N@20 & H@20 & N@10 & H@10 & N@20 & H@20 & N@10 & H@10 & N@20 & H@20 \\
    \midrule
    GRU4Rec & 0.0208 & 0.0412 & 0.0273 & 0.0670 & 0.0069 & 0.0146 & 0.0101 & 0.0274 & 0.0084 & 0.0174 & 0.0121 & 0.0325 & 0.0032 & 0.0066 & 0.0046 & 0.0123  \\
    w\textbackslash{} Ours & \textbf{0.0286} & \textbf{0.0546} & \textbf{0.0360} & \textbf{0.0842} & \textbf{0.0164} & \textbf{0.0307} & \textbf{0.0208} & \textbf{0.0483} & \textbf{0.0140} & \textbf{0.0289} & \textbf{0.0194} & \textbf{0.0502} & \textbf{0.0063} & \textbf{0.0128} & \textbf{0.0084} & \textbf{0.0214} \\
    Improve & 37.50\% & 32.52\% & 31.87\% & 25.67\% & 137.68\% & 110.27\% & 105.94\% & 76.28\% & 66.67\% & 66.09\% & 60.33\% & 54.46\% & 96.88\% & 93.94\% & 82.61\% & 73.98\% \\
    \midrule
    NextItNet & 0.0163 & 0.0326 & 0.0210 & 0.0511 & 0.0077 & 0.0154 & 0.0106 & 0.0272 & 0.0109 & 0.0222 & 0.0155 & 0.0406 & 0.0033 & 0.0070 & 0.0046 & 0.0125  \\
    w\textbackslash{} Ours & \textbf{0.0202} & \textbf{0.0365} & \textbf{0.0253} & \textbf{0.0589} & \textbf{0.0091} & \textbf{0.0193} & \textbf{0.0128} & \textbf{0.0320} & \textbf{0.0134} & \textbf{0.0282} & \textbf{0.0183} & \textbf{0.0465} & \textbf{0.0040} & \textbf{0.0087} & \textbf{0.0063} & \textbf{0.0162} \\
    Improve & 23.93\% & 11.96\% & 20.48\% & 15.26\% & 18.18\% & 25.32\% & 20.75\% & 17.65\% & 22.94\% & 27.03\% & 18.06\% & 14.53\% & 21.21\% & 24.29\% & 36.96\% & 29.60\% \\
    \midrule
    SASRec & 0.0338 & 0.0639 & 0.0413 & 0.0935 & 0.0174 & 0.0320 & 0.0214 & 0.0482 & 0.0136 & 0.0277 & 0.0180 & 0.0453 & 0.0078 & 0.0149 & 0.0100 & 0.0239  \\
    w\textbackslash{} Ours & \textbf{0.0455} & \textbf{0.0810} & \textbf{0.0539} & \textbf{0.1145} & \textbf{0.0242} & \textbf{0.0436} & \textbf{0.0294} & \textbf{0.0641} & \textbf{0.0164} & \textbf{0.0326} & \textbf{0.0216} & \textbf{0.0537} & \textbf{0.0128} & \textbf{0.0223} & \textbf{0.0154} & \textbf{0.0327} \\
    Improve & 34.62\% & 26.76\% & 30.51\% & 22.46\% & 39.08\% & 36.25\% & 37.38\% & 32.99\% & 20.59\% & 17.69\% & 20.00\% & 18.54\% & 64.10\% & 49.66\% & 54.00\% & 36.82\% \\
    \midrule
    FMLP-Rec & 0.0298 & 0.0563 & 0.0361 & 0.0814 & 0.0131 & 0.0255 & 0.0163 & 0.0383 & 0.0093 & 0.0195 & 0.0134 & 0.0357 & 0.0071 & 0.0134 & 0.0091 & 0.0215  \\
    w\textbackslash{} Ours & \textbf{0.0441} & \textbf{0.0767} & \textbf{0.0519} & \textbf{0.1076} & \textbf{0.0240} & \textbf{0.0432} & \textbf{0.0286} & \textbf{0.0615} & \textbf{0.0180} & \textbf{0.0347} & \textbf{0.0233} & \textbf{0.0559} & \textbf{0.0147} & \textbf{0.0245} & \textbf{0.0174} & \textbf{0.0353} \\
    Improve & 47.99\% & 36.23\% & 43.77\% & 32.19\% & 83.21\% & 69.41\% & 75.46\% & 60.57\% & 93.55\% & 77.95\% & 73.88\% & 56.58\% & 107.04\% & 82.84\% & 91.21\% & 64.19\% \\
    \bottomrule
    \end{tabular}}%
    \caption{Performance comparison of four backbone models and BASRec on four datasets. The `w/ Ours' represents adding our BASRec. All improvements are statistically significant, as determined by a paired t-test with $p \leq 0.05$.}
  \label{tab:main}%
\end{table*}}

\setlength{\tabcolsep}{1.5mm}{
\begin{table*}[!t]
  \centering
  \scalebox{0.80}{
    \begin{tabular}{c|cccccccc|cccccccc}
    \toprule
    BackBone & \multicolumn{8}{c|}{GRU4Rec} & \multicolumn{8}{c}{SASRec} \\
    \midrule
    \multirow{2}[2]{*}{Method} & \multicolumn{2}{c}{Beauty} & \multicolumn{2}{c}{Sports} & \multicolumn{2}{c}{Yelp} & \multicolumn{2}{c|}{Home} & \multicolumn{2}{c}{Beauty} & \multicolumn{2}{c}{Sports} & \multicolumn{2}{c}{Yelp} & \multicolumn{2}{c}{Home} \\
        & N@10 & H@10 & N@10 & H@10 & N@10 & H@10 & N@10 & H@10 & N@10 & H@10 & N@10 & H@10 & N@10 & H@10 & N@10 & H@10 \\
    \midrule
    Base & 0.0208 & 0.0412 & 0.0069 & 0.0146 & 0.0084 & 0.0174 & 0.0032 & 0.0066 & 0.0338 & 0.0639 & 0.0174 & 0.0320 & 0.0136 & 0.0277 & 0.0078 & 0.0149 \\
    \midrule
    Ran & 0.0212 & 0.0471 & 0.0083 & 0.0171 & 0.0087 & 0.0189 & 0.0036 & 0.0075 & 0.0285 & 0.0553 & 0.0186 & 0.0341 & 0.0162 & 0.0316 & 0.0086 & 0.0156 \\
    SW & 0.0192 & 0.0501 & 0.0082 & 0.0159 & 0.0090 & 0.0195 & 0.0040 & 0.0083 & 0.0270 & 0.0542 & 0.0198 & 0.0366 & 0.0137 & 0.0279 & 0.0089 & 0.0166 \\
    Ran-S & 0.0231 & 0.0520 & \underline{0.0103} & \underline{0.0207} & 0.0096 & 0.0210 & 0.0049 & 0.0099 & 0.0289 & 0.0563 & 0.0167 & 0.0300 & \textbf{0.0184} & \textbf{0.0364} & \underline{0.0109} & \underline{0.0208} \\
    CMR & 0.0225 & \textbf{0.0572} & 0.0095 & 0.0192 & \underline{0.0105} & 0.0209 & 0.0044 & 0.0094 & 0.0290 & 0.0562 & 0.0192 & 0.0374 & 0.0136 & 0.0278 & 0.0092 & 0.0167 \\
    CMRSI & \underline{0.0242} & 0.0555 & 0.0090 & 0.0183 & 0.0101 & \underline{0.0214} & \underline{0.0051} & \underline{0.0104} & \underline{0.0316} & \underline{0.0603} & \underline{0.0203} & \underline{0.0395} & 0.0156 & 0.0310 & 0.0099 & 0.0173 \\
    \midrule
    BASRec & \textbf{0.0286} & \underline{0.0546} & \textbf{0.0164} & \textbf{0.0307} & \textbf{0.0140} & \textbf{0.0289} & \textbf{0.0063} & \textbf{0.0128} & \textbf{0.0455} & \textbf{0.0810} & \textbf{0.0242} & \textbf{0.0436} & \underline{0.0164} & \underline{0.0326} & \textbf{0.0128} & \textbf{0.0223} \\
    \bottomrule
    \end{tabular}}%
    \caption{Performance comparison of heuristic augmentation methods and BASRec on four datasets. All improvements are statistically significant, as determined by a paired t-test with the second best result in each case $\left(p \leq 0.05\right)$.}
    \vspace{-1em}
  \label{tab:heuristic_aug}%
\end{table*}}%

\noindent \textbf{Evaluation Settings.} We adopt the leave-one-out strategy to partition each user’s item sequence into training, validation, and test sets. We rank the prediction over the whole item set rather than negative sampling, otherwise leading to biased discoveries \cite{krichene2020sampled}. The evaluation metrics include Hit Ratio@K (denoted by H@K), and Normalized Discounted Cumulative Gain@K (N@K). We report results with K $\in \{10, 20\}$. 

\subsection{Main Results with Various Backbone Models}
The experimental results of the original SR models and adding our BASRec are presented in Table \ref{tab:main}. We can observe that BASRec can significantly improve the performance of various types of SR models. Average performance gains on GRU4Rec, NextItNet, SASRec, and FMLPRec were 72.04\%, 21.76\%, 33.84\%, and 68.50\%, respectively. This result shows that the samples synthesized by our method can enhance the model's ability to learn user preferences further. Our approach achieves a win-win situation for both relevance and diversity, significantly improving the performance while ensuring plug-and-play generalization. Besides, the performance of the original model shows that SASRec outperforms the other models overall, demonstrating the power of the transformer in sequence modeling. With BASRec, the SASRec and FMLP-Rec are on par with each other. We believe that different models may have different underutilized performance potential. Our approach further exploits this potential through multiple mixup strategies.

\subsection{Comparison with Data Augmentation Methods}
We compare BASRec with different augmentation methods. For heuristic methods, we choose two representative models, SASRec and GRU4Rec, as the backbone network. For methods that require training, we chose the SASRec since it is available as a backbone network for all baselines.

\setlength{\tabcolsep}{1.25mm}{
\begin{table}[!t]
  \centering
  \scalebox{0.72}{
    \begin{tabular}{c|cccccccc}
    \toprule
    \multirow{2}[2]{*}{Method} & \multicolumn{2}{c}{Beauty} & \multicolumn{2}{c}{Sports} & \multicolumn{2}{c}{Yelp} & \multicolumn{2}{c}{Home} \\
    & N@10 & H@10 & N@10 & H@10 & N@10 & H@10 & N@10 & H@10 \\
    \midrule
    Base & 0.0338 & 0.0639 & 0.0174 & 0.0320 & 0.0136 & 0.0277 & 0.0078 & 0.0149 \\
    \midrule
    ASReP & 0.0351 & 0.0664 & 0.0195 & 0.0353 & 0.0162 & 0.0319 & 0.0099 & \underline{0.0184} \\
    DiffuASR & \underline{0.0372} & 0.0679 & 0.0202 & 0.0387 & 0.0150 & 0.0308 & 0.0105 & 0.0179 \\
    CL4SRec & 0.0366 & \underline{0.0686} & \underline{0.0221} & \underline{0.0412} & \textbf{0.0176} & \textbf{0.0355} & \underline{0.0119} & 0.0212 \\
    \midrule
    BASRec & \textbf{0.0455} & \textbf{0.0810} & \textbf{0.0242} & \textbf{0.0436} & \underline{0.0164} & \underline{0.0326} & \textbf{0.0128} & \textbf{0.0223} \\
    \bottomrule
    \end{tabular}}%
    \caption{Performance comparison of training-required methods and BASRec. The backbone network is SASRec.}
    \vspace{-1em}
  \label{tab:training_aug}%
\end{table}}%

\ \\ \noindent \textbf{Heuristic Methods.} Table \ref{tab:heuristic_aug} shows that BASRec outperforms existing heuristic augmentation methods in most cases. Our method generates new training samples by mixing representations in the representation space, which can better preserve the original sequence semantics and discover more cross-user preferences than existing methods. Among the baseline methods, CMRSI and Ran-S usually perform better. This suggests that well-designed data augmentation operators or augmentation using interactions in the original sequence are effective. In some cases, existing methods lead to model performance degradation. We believe this may be related to the fact that these methods introduce much noise into the augmented data, which interferes with model learning. In summary, it is essential to balance relevance and diversity in data augmentation, and favoring one side too much can lead to performance degradation.

\ \\ \noindent \textbf{Training-required Methods.}
Table \ref{tab:training_aug} shows the performance of BASRec compared to the training-required baselines. These methods usually contain auxiliary tasks or data generation modules that require training. BASRec achieves the best or second-best results without increasing model parameters. CL4SRec usually performs best among the baseline methods, indicating that contrastive learning can effectively mine preference information from sparse data.

\setlength{\tabcolsep}{1.25mm}{
\begin{table}[!t]
  \centering
  \scalebox{0.72}{
    \begin{tabular}{c|cccccccc}
    \toprule
    \multirow{2}[2]{*}{Method} & \multicolumn{2}{c}{Beauty} & \multicolumn{2}{c}{Sports} & \multicolumn{2}{c}{Yelp} & \multicolumn{2}{c}{Home} \\
    & N@10 & H@10 & N@10 & H@10 & N@10 & H@10 & N@10 & H@10 \\
    \midrule
    Base & 0.0338 & 0.0639 & 0.0174 & 0.0320 & 0.0136 & 0.0277 & 0.0078 & 0.0149 \\
    \midrule
    w/o SA & 0.0415 & 0.0734 & 0.0209 & 0.0374 & 0.0152 & 0.0295 & 0.0097 & 0.0184 \\
    w/o ALW & 0.0439 & 0.0778 & 0.0229 & 0.0415 & 0.0146 & 0.0297 & 0.0113 & 0.0201 \\
    w/o CA & 0.0397 & 0.0721 & 0.0193 & 0.0353 & 0.0145 & 0.0290 & 0.0098 & 0.0170 \\
    w/o NL & 0.0422 & 0.0752 & 0.0227 & 0.0412 & 0.0152 & 0.0300 & 0.0111 & 0.0197 \\
    w/o Two & 0.0416 & 0.0742 & 0.0231 & 0.0411 & 0.0151 & 0.0304 & 0.0117 & 0.0207 \\
    \midrule
    BASRec & \textbf{0.0455} & \textbf{0.0810} & \textbf{0.0242} & \textbf{0.0436} & \textbf{0.0164} & \textbf{0.0326} & \textbf{0.0128} & \textbf{0.0223} \\
    \bottomrule
    \end{tabular}}%
    \caption{Performance of different variants of BASRec. The backbone network is SASRec.}
    \vspace{-1em}
  \label{tab:ablation}%
\end{table}}%

\subsection{Ablation Study}
We conduct an ablation study to explore the effectiveness of various components in our method. We compare our BASRec with the following variants: 1) w/o SA: remove the Single Augmentation. 2) w/o ALW: remove the Adaptive Loss Weighting in Single Augmentation. 3) w/o Out: remove the Cross Augmentation. 4) w/o NL: remove the Nonlinear Mixup strategy in Cross Augmentation, i.e., perform general linear Mixup. 5) w/o Two: Use Eq. \ref{eq:main+aug} to jointly train the model from scratch without a two-stage training strategy.

The results are shown in Table \ref{tab:ablation}. The performances decrease significantly after removing either the Single Augmentation or the Cross Augmentation, suggesting that data augmented by both modules contributes to model training. When we replace Adaptive Loss Weighting with consistent weights, the model is unable to measure the difference between the enhanced data and the original data. Learning and distinguishing this difference can improve the model's performance and robustness. Nonlinear Mixup further extends the possibility of Cross Augmentation to generate augmented samples, and nonlinear combinations between different samples can help the model learn cross-preferences and fine-grained preferences. Besides, jointly training from scratch results in inferior performance compared to BASRec in four datasets, highlighting the significance of the two-stage training procedure. When two-stage training is used, the model can learn accurate representations in the first stage, while the second stage produces high-quality augmented representations by mixing these representations. If Mixup is performed at the beginning of training, inaccurate representations may interfere with each other, which is detrimental to model learning and convergence.

\section{Data Similarity Analysis}

\begin{table}[!t]
  \centering
  \scalebox{0.85}{
    \begin{tabular}{c|cccc|c}
    \toprule
    Method & Beauty & Sports & Yelp & Home & Average \\
    \midrule
    CMRSI & 0.5673 & 0.7273 & 0.6324 & 0.6884 & 0.6539 \\
    ASReP & 0.9585 & 0.9476 & 0.9307 & 0.9752 & 0.9530 \\
    CLS4Rec & 0.9612 & 0.9574 & 0.9580 & 0.9631 & 0.9599 \\
    \midrule
    BASRec-I & 0.9083 & 0.8954 & 0.8626 & 0.9522 & 0.9046 \\
    BASRec-O & 0.9239 & 0.9172 & 0.9003 & 0.9460 & 0.9219 \\
    BASRec & 0.9152 & 0.9076 & 0.8790 & 0.9485 & 0.9126 \\
    \bottomrule
    \end{tabular}}%
      \caption{Cosine similarity between the generated samples and the original samples. The backbone network is SASRec. The `BASRec-I' and `BASRec-O' represent only use Single-sequence augmentation and Cross-sequence augmentation, respectively.}
      \vspace{-1em}
  \label{tab:cos}%
\end{table}%

We calculated the cosine similarity between the samples generated by different data augmentation methods and the original samples. Since discrete interaction data cannot be computed directly, we compute the similarity of the final output representation of the model for all samples. We report the average similarity value throughout the training process and present the result in Table \ref{tab:cos}.

The table shows that the heuristic augmentation method CMRSI generates samples with relatively low similarity (i.e., lack of relevance), resulting in the loss of important preference knowledge and sequence semantics contained in the original samples. For training-required augmentation methods, ASReP employs a bi-directional Transformer to generating similar sequence data directly. The contrastive learning objective in CL4SRec draws close the distance between the original and generated samples. These methods result in high similarity between the augmented and original samples (i.e., lack of diversity). Our method generates samples with suitable similarity through a well-designed augmentation and fusion strategy. The new samples are diverse enough but retain the important semantics of the original data, balancing the relevance and diversity.

\section{Conclusion}
This paper introduces BASRec, a Balanced Data Augmentation Plugin for Sequential Recommendation, which aims to balance the relevance and diversity of augmented data. Our approach consists of Single-sequence Augmentation and Cross-sequence Augmentation. The former balances the two properties by heuristic operations with elaborate fusion and reweighting strategy. The latter fuses sequence representations from different users to generate samples that are diverse but retain important semantics of the original sequence. Extensive experiments demonstrate the superiority of BASRec. It improves the performance of various sequential models and outperforms existing methods. Our work emphasizes the importance of balancing relevance and diversity and demonstrates the great potential of augmentation in the representation space. For future work, we will further improve the method so that it can be integrated into other recommendation models. Moreover, we are interested in improving the simplicity of the method, e.g., by adaptively choosing operator rates and mixup weights.

\section{Acknowledgments}
This work is partially supported by the National Natural Science Foundation of China under Grant No. 62032013, No. 62102074, and the Science and technology projects in Liaoning Province (No. 2023JH3/10200005).

\bibliography{aaai25}

\newpage
\
\newpage

\appendix
\section{Appendix}
\subsection{Details for Implementations and Baselines}
All experiments are performed on a single NVIDIA RTX 3090Ti GPU with an Intel Core i7-12700 CPU and 32GB RAM. All models are implemented using the PyTorch framework version 1.12.1+cu116.

The baselines consist of three categories. The first category is general models to validate the effectiveness of BASRec. These models employ diverse architectures, including RNN, CNN, Transformer, and MLP. The second category is heuristic augmentation methods. The third category is training-required augmentation models.

\subsubsection{General Models}
\begin{itemize}
    \item \textbf{GRU4Rec} \cite{hidasi2015session}: This model leverages gated recurrent
units to capture behavioral patterns.
    \item \textbf{NextItNet} \cite{yuan2019simple}: This model combines masked filters with
1D dilated convolutions to model the long-range dependencies. 
    \item \textbf{SASRec} \cite{kang2018self}: It adopts the multi-head self-attention mechanism to perform sequential recommendation.
    \item \textbf{FMLPRec} \cite{zhou2022filter}: It is an all-MLP model with learnable filters for sequential recommendation.
\end{itemize}

\subsubsection{Heuristic Methods}
\begin{itemize}
    \item \textbf{Random (Ran)} \cite{liu2023diffusion}: This method augments each sequence by randomly selecting items from the whole item set
    \item \textbf{Random-seq (Ran-S)} \cite{liu2023diffusion}: It selects items from the original sequence randomly as the augmentation items.
    \item \textbf{Slide Windows (SW)} \cite{tang2018personalized}: It adopts slide windows to intercept
multiple new subsequences from the original sequence.
    \item \textbf{CMR} \cite{xie2022contrastive}: This work proposes three sequence data augmentation operators with contrastive learning. In this category, we only use the three operators, including Crop, Mask, and Reorder.
    \item \textbf{CMRSI} \cite{liu2021contrastive}: Based on CMR, this work proposes two informative augmentation operators with contrastive learning. We adopt five operators including Crop, Mask, Reorder, Substitute, and Insert.
\end{itemize}

\subsubsection{Training-required Methods}
\begin{itemize}
    \item \textbf{ASReP} \cite{jiang2021sequential}: This method employs a reversely pre-trained transformer to generate pseudo-prior items for short sequences. Then, fine-tune the pre-trained transformer to predict the next item.
    \item \textbf{DiffuASR} \cite{liu2023diffusion}:  It adopts the diffusion model for sequence generation. Besides, two guide strategies are designed to control the model to generate the items corresponding to the raw data.
    \item \textbf{CL4SRec} \cite{xie2022contrastive}:  This method leverages random data augmentation and utilizes contrastive learning to extract self-supervised signals from the original and augmented data.
\end{itemize}

We do not include methods ReDA \cite{bian2022relevant}, CASR \cite{wang2021counterfactual}, and L2Aug \cite{wang2022learning} since they do not provide open-source codes for reliable reproduction (empty code repository or no available code repository links).

\begin{figure}[!t]
	\centering
	\includegraphics[scale=0.38]{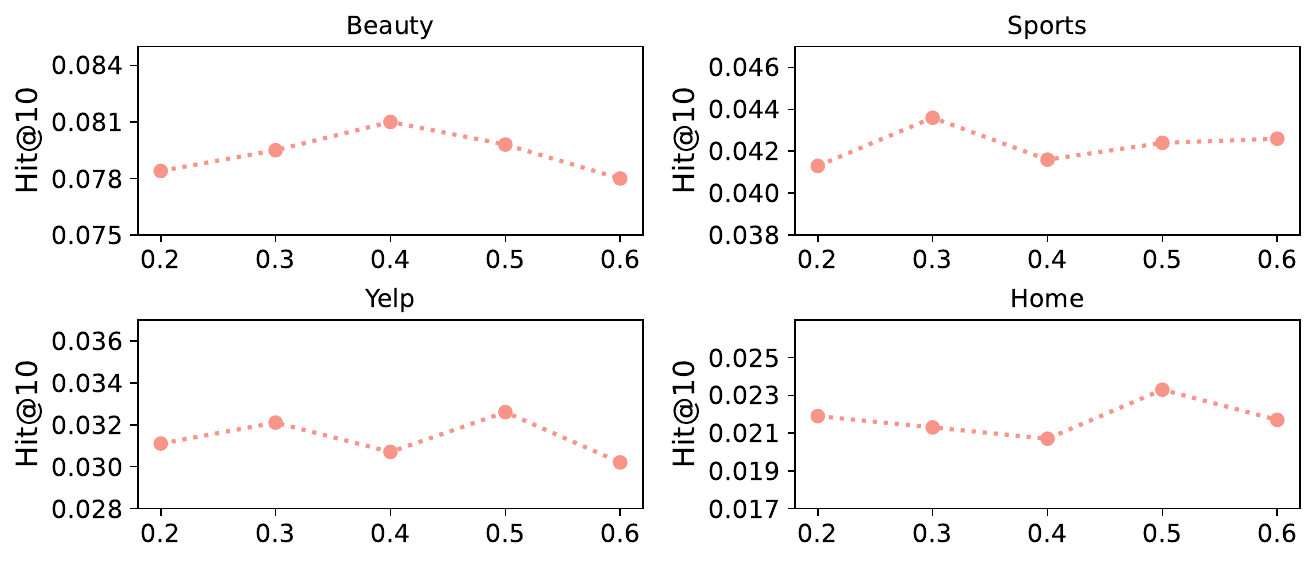}
	\caption{Sensitivity analysis of parameter $\alpha$.}
	\label{fig:beta}
\end{figure}

\begin{figure}[!t]
	\centering
	\includegraphics[scale=0.38]{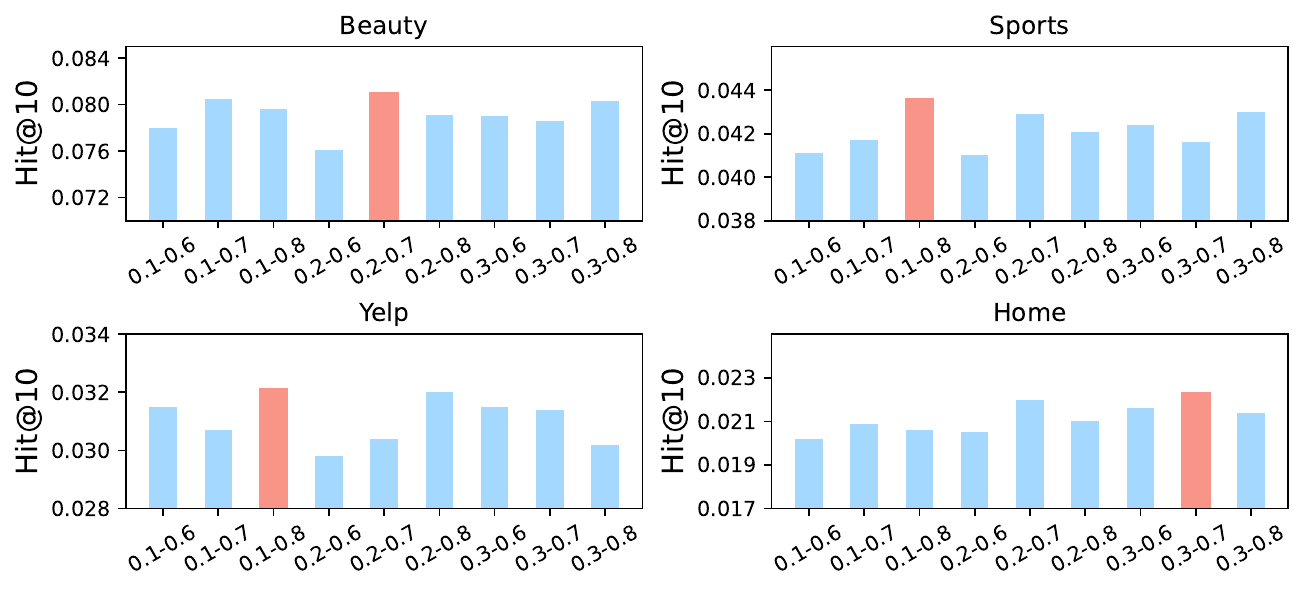}
	\caption{Sensitivity analysis of parameters $a$ and $b$. For example, `$0.1-0.6$' represents $a=0.1$ and $b=0.6$. We highlight the best result with a different color.}
	\label{fig:ab}
\end{figure}

\subsection{Hyper-parameter Analysis}
We further investigate the hyper-parameter $\alpha$ of beta distribution and $a,b$ of the operators. The results are illustrated in Figure \ref{fig:beta} and \ref{fig:ab}. In our approach, all beta distributions share the same $\alpha$, and the two operators share the same $a$ and $b$.

For $\alpha$, we find that optimal values tend to arise in $\{0.3,0.4,0.5\}$, with both larger and smaller values leading to degradation of model performance. The optimal $\alpha$ for the four datasets is $0.4, 0.3, 0.5, 0.5,$ respectively. For the $a$ and $b$ of operators, we observe that the optimal sequence operation ranges are concentrated in the middle. In other words, neither too many nor too few changes can be made to the original sequence. Also, a narrower range, such as only 0.4 for the `$0.2-0.6$', can lead to poorer performance. Overall, the choice of $a$ and $b$ needs to avoid values that are too small and too large but also allow sufficient room for choice between $a$ and $b$. This finding corroborates the need to balance relevance and diversity during data augmentation.

\end{document}